# Learn Like Feynman: Developing and Testing an AI-Driven Feynman Bot

Akshaya Rajesh
*Singapore University of Technology and Design*
Singapore
akshaya_rajesh@alumni.sutd.edu.sg

Sumbul Khan
*Science, Mathematics and Technology cluster*
*Singapore University of Technology and Design*
Singapore
sumbul_khan@sutd.edu.sg

*Abstract*— The Feynman learning technique is an active learning strategy that helps learners simplify complex information through student-led teaching and discussion. In this paper, we present the development and usability testing of the Feynman Bot, which uses the Feynman technique to assist self-regulated learners who lack peer or instructor support. The Bot embodies the Feynman learning technique by encouraging learners to discuss their lecture material in a question-answer-driven discussion format. The Feynman Bot was developed using a large language model with Langchain in a Retrieval-Augmented-Generation framework to leverage the reasoning capability required to generate effective discussion-oriented questions. To test the Feynman bot, a controlled experiment was conducted over three days with fourteen participants. Formative and summative assessments were conducted, followed by a self-efficacy survey. We found that participants who used the Feynman Bot experienced higher learning gains than the Passive Learners' group. Moreover, Feynman Bot Learners' had a higher level of comfort with the subject after using the bot. We also found typing to be the preferred input modality method over speech, when interacting with the bot. The high learning gains and improved confidence with study material brought about by the Feynman Bot makes it a promising tool for self-regulated learners.

## I. INTRODUCTION

Over the past few years, generative artificial intelligence, specifically large-language models, have been capitalized by learners and educators for many learning tasks, such as writing, proofing, and summarizing. Many applications have recognized the potential of using generative AI to promote active learning. Techniques such as question-answering and flashcard-based studying have been digitized to become easier to perform using apps such as AnkiBrain[1], Revision.ai [2]. Self-studying applications such as Anki [3] have helped learners use spaced-repetition techniques for knowledge retention, while ChatGPT [4] is used by students to explore noted by asking questions and getting direct relevant answers. These applications help learners explore complex topics faster, cutting down study preparation time and using active learning techniques.

Active learning focuses on student-centric learning techniques to increase understanding and performance by enhancing learners' engagement. It encompasses a wide array of methods of varying levels of engagement that learners can use to achieve a deep level of knowledge and comprehension of information.

Several papers highlight the ability of active learning to yield better scoring results and lower failure rates in learners, while also considering the link between student perceptions of these techniques and the quality of learning they yield. [5-6]. An extensive meta-analysis of studies on examination scores and failure rates of learners using active learning vs passive learning in STEM carried out by study [5], not only highlights the 6% increase in examination scores of active learning learners, but also mentions increased perception of usefulness as the level of activity in the method of learning increased [5]. However, there are several roadblocks that inhibit the use of active learning approaches. First, studies have found low self-efficacy, shyness, and fear of leaving the comfort zone as student barriers to active learning [7]. Second, studies have cited that there are administrative roadblocks for active learning to be implemented regularly in a classroom setting [7]. Furthermore, there is a lack of comprehension of active learning methodologies which results in learners defaulting to passive learning techniques.

Self-regulated learners learn outside the classroom setting, hence circumventing administrative roadblocks that may limit their learning techniques. However, they lack peer or instructor support to try out active learning techniques such as the Feynman Technique.

We posit that there is an opportunity to incorporate an active learning approach like Feynman technique in a self-study or instructor-independent setting, through online platform or application-based implementation. This alternative has the potential to not only shorten the time span of active learning implementation, but also provide higher benefits than when used an instructor-dependent environment [5]. Affective barriers to active learning such as lack of confidence, fear of being incorrect and low participation skills can be overcome by first having the student build up confidence and comfort in the self-study setting, before implementing the learning method in a collaborative space [7-9].

While current technologies such as Anki [3] and Revision.ai [2] help self-regulated learners in memorization, flashcard preparation and question-answers, none of them encourage a heutagogical teaching of the topic that encourages the student to activate their higher order reasoning. There is a need for an accessible Bot to help learners learn and understand concepts of topics relevant to their level of study using effective active learning techniques. This Bot needs to use study time to increase the learner's comfort with the subject and their learning gains through constant high engagement with the material.

To cover this gap we developed the Feynman Bot, an educational Bot that adapts the Feynman technique to simulate a learning-by-teaching scenario using artificial intelligence to interact with learners. The bot's novelty lies in its two-way conversational ability where the student actively teaches concepts and the Bot asks questions, points out inaccuracies, and questions deeper aspects of concepts through scenario-based discussions. In this paper, we present the development of Feynman Bot as an app targeting self-regulated learners to use high engagement active learning for study. We test the Feynman Bot through a remote unmoderated usability test conducted over three days with 14 participants.

The main contributions of this study are:

1. We present the development of a learning app that embodies the Feynman technique, specially targeted towards self-regulated learners,

2. We present findings about the effectiveness of the app on learning gains and student comfort with the study material

3. We present findings about preference of input modality for Feynman Bot

II. BACKGROUND

*A. Active Learning*

Studies performed on active learning have established its importance and effectiveness in increasing student comprehension and performance in topics across STEM and biological sciences [5-6]. Active learning techniques, rooted in urging learners to engage with material better by thinking, discussing, investigating, and creating, are shown to increase student performance in examinations and reduce the odds of failing [6,11]. It is targeted towards increasing student interaction through techniques that primarily fall under five pedagogies: problem-based learning, discovery-based learning, research-based learning, project-based learning, and case-based learning [11-12]. The learner-centered approach achieves higher student engagement by emphasizing critical thinking, discussion, and information analysis [5]. A large corpus of studies performed to investigate the effectiveness of active learning techniques have proven an increase in student performance at different grade levels [5]. Learners are seen to make enhanced efforts when using this method, leading to low-performing learners moving to a higher grade level [5,7]. This increase in testing results is attributed to the higher levels of engagement in the same study period [13]. Numerous studies have shown that the attention span of learners during a lecture is roughly 15 mins. Passive learning techniques, such as watching lectures or reading, result in information deliverance beyond the student's attention span. Engagement during active learning holds student attention resulting in better learning outcomes [13-14]. Moreover, active learning reveals learners' fundamental misconceptions and requires their clarification, which is an essential element of effective teaching [15].

*B. The Feynman Technique*

The Feynman Technique, which falls under active learning, is a form of heutagogical learning pioneered by renowned physicist Richard Feynman, that encourages learners to teach topics to peers and to have discussions through which loopholes in knowledge can be exposed [16]. The technique involves deep dive questioning of explanations and discussing the application of concepts to scenarios to uncover gaps in the learner's understanding. The learner can clarify these gaps by going back to the source material, and then continue the teaching method to uncover more issues they may face with the topic.

The Feynman Technique has been implemented in both classroom and remote instructor-independent studies and has resulted in higher learning gains [16]. The technique also results in an improvement in communication skills through student participation [16-17].

*C. Existing AI based tools for Active Learning*

The employment of active learning during self-study has been made more accessible by several AI-driven applications, that cut down on preparations required for active learning by generating question-answers, organizing spaced-repetition revision schedules etc. Tools such as ChatGPT [4] also help learners go through notes on complex topics faster by providing a chat interface with documents. Collectively, a combination of these applications can help learners use active learning on complex topics, with little preparatory work. However, there is no application that individually addresses all requirements.

Although applications that employ active learning methods such as question-answers, flashcards and spaced repetition are used widely, they target a student audience that needs to memorize their study material. The engagement aspect of active learning has not been utilized for deeper comprehension. An application that can stimulate higher order thinking through dynamic two-way discussion, while being easy to use is still lacking. Facilitation of higher-order thinking requires a student-centered technique with a higher level of active engagement than provided by simple flashcards. We also want to address the limitation posed by a lack of peers in a self-study setting, as the Feynman Technique is predicated on having a partner ask deeper questions and ensure answers given are accurate.

This paper aims to address the gap in discussion-based learning technology by incorporating the Feynman technique in a manner that overcomes the limitation of self-studying on its use.

## III. METHODOLOGY

### A. Development of app

The Feynman Bot is a generative AI Bot that stimulates context specific discussions to help students learn using the Feynman technique. We employed large language models in a retrieval-augmented-generation (RAG) framework, to make the Bot capable of understanding natural language in the context of the learner's subject and level. This also ensures factual correctness in learner explanations without going out of their scope of learning. From an architectural perspective, the RAG framework allows scalability and flexibility in the size and type of documents provided by the user.

The main components of the Feynman Bot application's architecture are presented in Fig. 1.

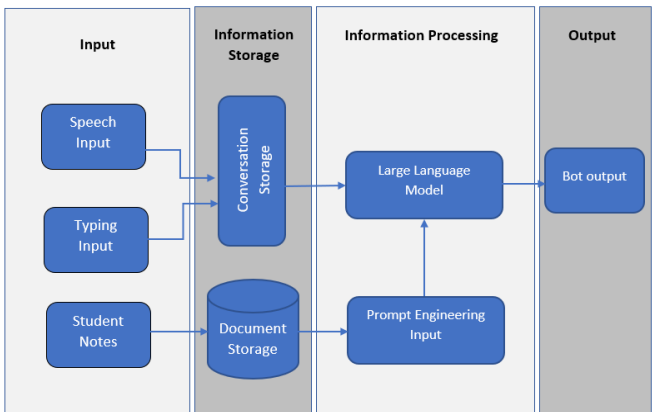

Fig. 1: Feynman Bot architecture diagram

### B. Key features of app

Feynman Bot incorporates the following key features to facilitate student-led conversation:

1. *Large Language Model*: The Feynman Bot employs a large language model that ingests student notes and prompt instructions as inputs. The output is a discussion-based dialogue that either begins a conversation or builds on the student's previous answer. We capitalize the ability of large language models to understand the learner's intent and follow their line of thought. It uses this understanding to generate dialogue and embody the Feynman technique.
2. *Prompt Engineering*: The Feynman Bot uses prompt engineering to ensure the model uses the Feynman Method while adhering to the context (study materials) provided. Prompt engineering provides a set of instructions to the large language model to control its responses and knowledge scope. This not only ensures the Bot uses the Feynman Technique, but also limits the scope of knowledge learnt to that provided by the user as input notes, thus preventing hallucinations.
3. *Student Notes Document Upload*: The model ingests student notes or chapters to use as context. This allows the Bot to focus the knowledge space and subject level of the discussion,
4. *Information Retrieval*: The student notes or documents are stored temporarily for the conversation. Relevant information is retrieved from them according to the conversation's direction. The Bot uses the information retrieved to augment the generation of its responses and questions, a process known as retrieval augmented generation (RAG).
5. *Text Input*: The Feynman Bot allows learners to type their responses and explanations while chatting with the bot. This is provided for learners who prefer to deliberate and phrase answers precisely before entering it as their response.
6. *Speech Input*: A speech input capability was built into the Feynman Bot to allow learners to speak out their explanations and answers during the conversation. As the learner talks through their line of thought, the Bot understands their logic. This gives the Feynman Bot more information to better pinpoint areas of misconception or loopholes in the student's knowledge and guide the conversation toward these areas.

An example of a conversation of a learning with the Feynman Bot is presented in Fig 2.

| Entity | Dialogue |
|---|---|
| Bot | *Let's explore the concept of cost-benefit analysis within utilitarianism. Can you tell me about how cost-benefit analysis is used and a case-study reflecting the use in a real-world scenario?* |
| FB9 | *The Philip Morris case study exemplifies the way the government could place a dollar value to the profits and losses of banning smoking, therefore resulting in a financial-oriented effects of the potential ban for the government. Cost-benefit analysis was used in terms of measuring loss of profits in the sales of cigarettes, as well as increased costs invested into the longer lifespans and healthcare of non-smoking citizens. In the end, the verdict based on the potential gain in profits of smokers dying early was that allowing smoking was more financially beneficial to the country. The extra tax revenue generated from cigarette sales, reduced need for pensions due to shortened lifespans, and additionally more freed up housing due to citizens dying earlier overcame the financial benefits of citizen's increased healthcare savings from not needing treatment for smoking-related diseases like lung cancer. This resulted in the conclusion that the overall happiness financially for the government was greater when allowing smoking, as opposed to banning it.* |
| Bot | *Great explanation! Could you link this back to the utilitarian logic of maximizing utility?* |
| FB9 | *The utilitarian logic overlooks the other sufferings of citizens that were non-fiscal and immaterial. It failed to place value on quality of life of citizens in terms of their health despite being alive, air quality, and that shortened lifespans would negatively impact the families of those who were impacted by smoking health-wise. It also failed to account for the loss in contribution of said individuals due to ailing health caused by smoking and the shortened lifespan* |
| Bot | *That's great. This was a good overview on cost-benefit analysis, and it's use in a case-study.* |

Fig. 2: Example of conversation of a learner with the Feynman bot

### C. Experiment setup and data collection

We conducted a controlled experiment to test whether the Feynman Bot can increase learning gains and learner comfort with the study material, as compared to conventional passive study techniques in a self-regulated, instructor-independent setting.

The experiment was conducted individually with 14 working adults (9 females and 5 males, ranging between the ages of 21-29). Participants' educational backgrounds ranged from engineering (8), risk analysis (2), accounting (2), psychology (1) and education (1). None of the participants had any exposure to law or philosophy.

All participants were asked to enroll in the edX online course "Justice by Harvard," [18], a course that explores critical, classical, and contemporary theories of justice with discussions on present-day applications. Two lectures from the online course were selected to be the focus of the study: Lecture 1 "*Utilitarianism: Jeremy Bentham*" and Lecture 2, "*Utilitarianism: J.S.Mills* " by Michael J. Sandel. Each lecture covered the topic of utilitarianism but approached it from a different perspective. This variation allowed students to compare and contrast the lecture content, facilitating deeper discussions with the bot on the subject. The subject of study did not overlap with the education or job scope of any of the participants, making it an unbiased topic to test participants' learning in the experiment.

A random split was used to divide the participants into two groups: the Passive Learners group (control group, referred as CG) and the Feynman Bot Learners group (experimental group, referred as FB). The Passive Learners group was instructed to learn the material only by watching the lecture and reading notes, while the Feynman Bot Learners group was instructed to watch the lecture, immediately followed by a discussion with the Feynman Bot.

The experiment was conducted over 3 days, with Days 1 and 2 focusing on learning two lectures from the given online course, and Day 3 focusing on reviewing both lectures and completing the summative assessment.

A formative assessment was conducted to gauge participants' baseline level of knowledge on the subject prior to the assigned learning tasks. The formative assessment consisted of questions from the two lectures generated by Quizbot.ai [19], reviewed down and selected by the authors. This included 10 MCQ questions and 2 open-ended questions.

In the first segment on day 1, the participants were asked to watch Lecture 1 (*"Utilitarianism: Jeremy Bentham"* led by Michael J. Sandel [18]).

In the second segment, the participants were asked to review the material. Passive Learners could review the material by rewatching the lecture, taking notes or re-reading notes. On the other hand, the Feynman Bot Learners were asked to upload the transcript of the lecture, as the input document (available as part of the course) to the Feynman Bot and have a 25-minute conversation with the Bot on the topic. The learners were instructed to either start the conversation by using the recorder button to speak to the bot, or by typing a starting prompt:

User: *Hi, this is a transcript of a lecture from a Harvard Justice module I am taking on Utilitarianism, can you help me review this so I understand it completely and can tackle scenario-based questions that may pop up in my exam*

A video tutorial was provided illustrating steps for using the user interface of the bot.

On day 2, participants were asked to watch the next lecture ("*Utilitarianism: J.S.Mill*" led by Michael J. Sandel [18]). After watching the lecture once, the Passive Learners were asked to use the remaining session to rewatch the lecture or take notes. The Feynman Bot Learners uploaded the transcript of the lecture and conversed with the Bot for 25 minutes.

On day 3, both groups were asked to review the material of the two lectures for 25 minutes. Passive Learners could read notes or watch the lecture, while Feynman Bot Learners uploaded both lecture transcripts and conversed with the bot.

A summative assessment was conducted for both participant groups. The summative assessment comprised of (1) Ten MCQ type questions (2) Two open-ended questions. Both categories of questions were generated by uploading the same two lecture materials into Quizbot.ai [19], by first generating 30 questions, followed by review, and down selection by the authors.

A self-efficacy survey was conducted with the Feynman Bot Learners group. The survey comprised of five Likert scale questions pertaining to learners' comfort with the study material and with using the Feynman Bot. A summary of the two learning groups' learning tasks is presented in Fig. 3.

| | Day 1 | | Day 2 | | Day 3 | |
|---|---|---|---|---|---|---|
| | LEARN LECTURE 1 | REVIEW LECTURE 1 | LEARN LECTURE 2 | REVIEW LECTURE 2 | REVIEW BOTH LECTURES | TEST |
| | Segment 1 (25 mins) | Segment 2 (25 mins) | Segment 1 (25 mins) | Segment 2 (25 mins) | Segment 1 (25 mins) | Segment 2 (25 mins) |
| Passive Learners | Complete Formative Assessment | Rewatch lecture or read notes | Watch the first Lecture on Utilitarianism: J.S. Mill | Rewatch lecture or read notes | Review contents of lecture through reading notes or rewatching lecture video | Complete summative assessments |
| Feynman Bot Learners | Watch first Lecture on Utilitarianism: Jeremy Bentham | Upload transcript of lecture on Feynman Bot and have conversation with the Feynman Bot | | Upload the transcript of the lecture on the Feynman Bot and have a conversation with the Feynman Bot | Upload transcript of the two lectures and converse with the Feynman Bot to review material | |

Fig. 3: Summary of learning tasks of the Passive Learners and the Feynman Bot Learners

*D. Data Analysis*

We analyzed the impact of the Feynman Bot in two areas:

(1) *Learning gains,* which is the improvement in learners' knowledge from intervention. We analyzed accuracy of responses in MCQs. We computed learning gain as the difference in the scores of the summative assessment and the formative assessment as shown here:

$$Learning\ Gain = Summative\ Score - Formative\ Score \qquad (1)$$

As the formative assessment and the summative assessment comprised 10 MCQs each, the maximum possible learning gain score was 10, and the minimum possible score was -10. Accuracy of responses in open-ended questions was analyzed using thematic analysis techniques.

(2) *Learners' self-efficacy* was measured through a 5-point Likert scale and was analyzed using statistical techniques.

IV. RESULT AND FINDINGS

All the participants successfully completed the learning tasks in the experiment with no dropouts. Participants also completed all questions in the formative and summative assessments. In total, 28 responses were recorded from the formative and summative assessments.

### A. Learning gain

The results showed that the mean summative score was greater for Feynman Bot Learners (M=6.8, SD=1.3) than the Passive Learners (M=6.5, SD=1.9). We found that the Feynman Bot group experienced a higher mean learning gain as compared to the Passive Learners while studying the same material for the same length of time (CG= 2.17, FB =3.83). The standard deviation of the learning gain score of the Feynman Bot Learners group, as indicated in Fig. 4, was also lower than that of the Passive Learners' group (Passive Learners' group = 1.94, Feynman Bot group= 1.32), indicating a more consistent improvement by those who utilized the Feynman Bot for learning than Passive Learners.

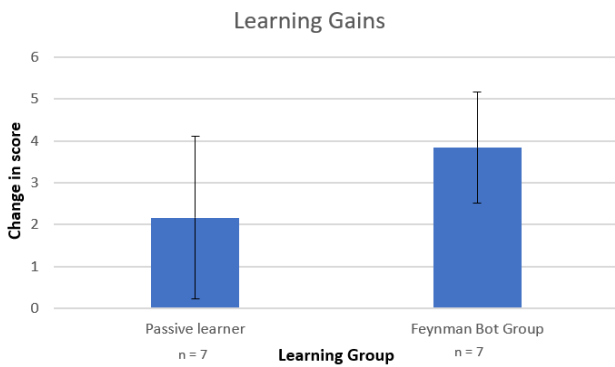

Fig. 4: Mean Learning Gains

Participants' individual learning gain scores are presented in Fig. 5. A noteworthy observation is that the learning gains exhibits a higher variation in length for members of the Passive Learners' group, including a member with negative improvement (indicating they did better in the formative than summative assessment). Such a large variation supports the standard deviation measure's conclusion on lower score consistency among Passive Learners group members.

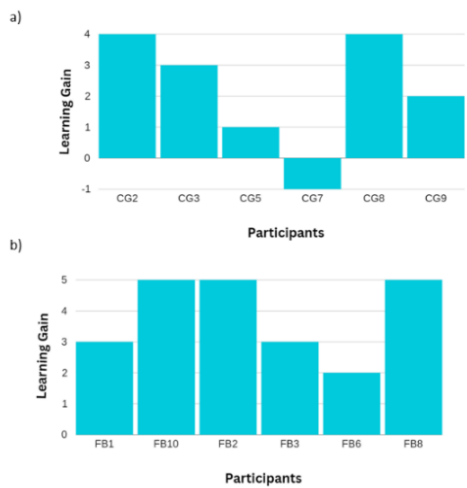

Fig. 5: Learning gain for each participant. (a) CG = Passive Learners' group (b) FB = Feynman Bot group (Outliers removed)

All the participants from both groups answered the open-ended questions correctly. A noteworthy observation was made in the style of answering of the open-ended questions in the two groups. Most participants in the Feynman Bot Learners group included scenario or case-study based evidence as further explanations to their answers. Meanwhile every member of the Passive Learners provided short answers, without any explanations. An example of participants' answers from each group can be seen in Fig. 6.

| Question | *"Discuss the objections to utilitarianism, particularly in terms of its impact on minority rights"*. |
|---|---|
| Feynman Bot user answer | *"One of Mill's argument against Bentham's theorem on having cost-benefit analyses done was on the point of its impact on minority rights. When we seek to maximize pleasure against pain of a collective group, a decision which would silence the voice of the minorities and sacrifice them may be made. He argues that such sacrifices should not be made, even at the cost of not maximizing utility for the collective group.*<br><br>*An example of cases where minority rights was brought up during the lecture, where the Romans would put Christians in the Colosseum to fight for their lives. While the utility (their life) of the Christians would be sacrificed for the utility (enjoyment) of the Romans, such a case puts the minority at the backseat, and neglected.*<br><br>*As such, Mills proposed that due to such cases, the maximum collective utility, in the case where there may be an indefinite number of Romans, such that the collective utility they would gain from such a showing would outweigh the lives of the Christians, should be avoided for the benefit of the minority."* (FB3) |
| Passive learner answer | *"Utilitarianism aims to make to make as many people as happy as possible, which can lead to minority rights being dismissed for the greater good".* (CG5) |

Fig. 6: Example open-ended responses

### B. Measurement of Self efficacy

We found that the Feynman Bot was regarded as a more effective mode of studying compared to usual passive study techniques. Over 80% of the Feynman Bot Learners group participants indicating they preferred using it to read notes or rewatching lecture materials repeatedly for study. All participants who used the Bot agreed that it increased their level of comfort with the study material enough to discuss the subject material with other peers or learners who took the same course.

Over 57% of the Feynman Bot Learners felt they were comfortable enough with the study material to teach or converse about it with new learners. The Bot was viewed as a useful study method for new topics according to 85.7% of participants.

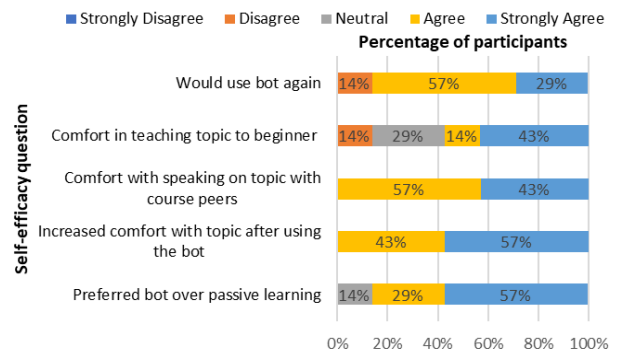

Fig. 7: Self-efficacy survey results

A final interesting finding, conveyed as feedback to researchers, was the preference for most Feynman Bot Learners to use chat like typing as an input modality as compared to speaking with the Bot directly.

## V. DISCUSSION

This study focused on the development and testing of the Feynman Bot, a context flexible bot that embodies the Feynman technique to help self-regulated learners understand and learn material through discussion. We tested the impact of the Bot on learning by conducting a controlled experiment with learners studying an online course. Formative and summative assessments of the Passive Learners' group and Feynman Bot group were conducted to compare the learning gain of each group. A self-efficacy survey taken after the study helped in determining the impact of the Bot on students' comfort with the study material after the study sessions.

Through the study conducted with 14 participants we found that learners who used the Bot experienced a higher learning gain than those who did not. Moreover, Feynman Bot Learners demonstrated a higher level of comfort and confidence with discussion on the thick theoretical study material. Lastly, we found that learners have a greater preference for typing as the input modality as opposed to speech while using the Bot for discussing a given topic.

### A. Impact on Learning Gain

A comparison of the two groups' learning gain showed that Feynman Bot users achieved a higher gain with smaller standard deviation in their scores than the Passive Learners Group.

The high variation in the scores of the Passive Learners' group could be caused due to student specific factors that impact performance such as communication, learning facilities, guidance etc. [10]. This causes some participants to standout with high improvement, such as student CG8 in Fig. 5, while students like CG7 demonstrate negative improvement. The Feynman Bot uses continuous engagement to reduce the effect of these student-specific factors by ensuring consistent effort and engagement across all learners. This subsequently translates to a lower standard deviation in the improvement scores. Moreover, this low variation is centered around a higher mean improvement score in the Feynman Bot group than the Passive Learners' group, meaning more Feynman Bot learners are performing at a consistently higher level compared to passive learners.

### B. Impact on Open-ended answer style

A key difference noted between the Feynman Bot Learners and Passive Learners' group was the elaboration provided while answering open-ended questions. Most of the participants in the Feynman Bot Learners group not only provided the correct answer to the question but elaborated with case-study based explanation from the lecture, as opposed to direct theoretical answers from the Passive Learners group. Their tendency to do this, without explicit instructions provided in the question, not only highlights their recall of the material, but their ability to apply discussion-based comprehension to exam questions. They could explain the concept in simple terms through an example (like the Feynman technique) to make the original theory more comprehensible and supported. This result supports Feynman bot's ability to increase student recall, comprehension, and comfort with the subject.

### C. Effect on student comfort and confidence

Over 80% of participants indicated that they would use the Bot again for a new topic. This suggests that the Bot helps in lowering barriers to integrating active learning into their self-regulated studying. The Bot overcomes the self-efficacy challenges learners have faced by implementing active learning techniques, increasing their confidence to not only speak about the subject matter with other peers in the course, but to teach it those who have not studied the subject before. The higher learning gain, improvement in student confidence and high acceptance of the Bot makes it promising for self-regulated learners. While the Feynman Bot method of learning was agreeable to 86% of users, 14% of the participants mentioned they would not use the bot again. This could most likely be attributed to previously mentioned student barriers to using active learning techniques such as low self-efficacy or discomfort with continuous high-engagement learning.

### D. Preference of input modality

The Feynman Bot application takes in the user input for the conversation in 2 modes: speech input and typing. An interesting finding from student participant feedback, was the higher preference for using the typing mode of input over speech input. Given the discussion-oriented nature of the conversation and the focus on using the Feynman technique for explanations, speech input was originally hypothesized as the preferred form of input. However, this hypothesis was disproven, as many participants felt that typing allowed them to articulate their thoughts and explanations better than if they used speech input. Some also preferred typing since it could be used in various settings (say learning in a cafe) compared to speech which could be impacted by the user's surrounding.

The results on learning gains from this study are in line with higher learning gains exhibited by students who use the Feynman Technique [1], indicating that the implementation of the technique is successful. Even in this exploratory study, our learning gains are comparable to the 6% increase in student performance mentioned in [16].

The Feynman Bot is a context flexible technology that ingests material the student is intent on learning and provides prompts to stimulate in-depth conversations, urging learners to think on core concepts of the subject and their influence at a higher level. It is designed to be used by self-regulated learners of different subject levels for a wide range of contexts. This makes its use cases span a wide range of demographics, from young school children to university students to working adults.

Previous studies have reported the gamification of active learning through application-development, yielding an 8%

learning gain in students [20]. However, these studies focus more on using student correctness tracking to order questions based on difficulty. The conversational working principle of the Feynman Bot is grounded in the pedagogy of the Feynman Technique, making its impact on learning gain high and consistent among all students.

## VI. Conclusion

This paper's objective was to present the development of the Feynman Bot and investigate its impact on student learning and comfort. The Feynman Bot is an educational Bot that embodies the Feynman technique to assist learners in self-regulated learning settings. It employs a two-way conversational capability to facilitate student-led dialogue, to improve their learning gain and comfort with their study material. The Bot employs large language models adjusted with prompt engineering to embody the Feynman technique through discussion. It ingests student study materials including lecture notes, transcripts, textbook chapters, and question-answers as input and formulates conversational prompts or questions to target the learners' learning objectives. The incorporation of these features into the Feynman Bot makes it a context flexible learning Bot that embodies the Feynman technique for higher engagement of self-regulated learners with their study material.

### A. Limitations and Future work

This study has the following limitations: (1) The study was performed on a small sample size. Although the results from this study are promising, it is necessary for this Bot to be tested on larger sample size. (2) The Feynman Bot is successfully implemented for use in theoretic subjects such as philosophy, law, biology that involve description of facts and concepts using language. The Bot needs to be further developed to expand its capabilities for subjects such as Mathematics, Chemistry and Physics that involve calculations, symbolic notation and numerical problem-solving. (3) Although the speech input capability in the current implementation can parse narrated calculations or problem-solving approaches, there needs to be a more accurate method of tracking the user's approach to problem solving. These limitations can be overcome through a further development of Feynman bot's architecture and a larger experiment, both of which will be detailed in future studies.

The successful implementation of active learning technology and its targeted benefits for self-regulated learners has been demonstrated by the Feynman Bot. Future developments extend the use of the bot in a classroom setting so that it can be used by learners and teachers to explore subjects in a more exciting and communicative fashion.

The study presented in this paper demonstrates the use case of working adults taking an online course and successfully boosts their learning gain and comfort with the material. Feynman Bot learners display more consistent learning gains and higher recall of the material when answering open-ended questions. Future studies using the bot would focus on further validating these results with a larger group of participants over a longer period to measure the long-term effects of using the Feynman Bot as compared to passive learning. We would also focus on developing the bot's accuracy in ensuring high-engagement study session for calculation-based subjects.

In this paper, we have designed and implemented an innovative Bot that successfully maintains high engagement study session through a two-way conversation on the material. The bot is a powerful solution for self-regulated instructor independent learners who want to use active learning techniques for deep understanding of their own study materials. Our study displays the success of the bot's implementation in increasing the learning gains and student comfort.